\documentclass[subscriptcorrection,upint,varvw,barcolor=black,mathalfa=cal=euler,balance,hyphenate,french,pdf-a]{article} %
\usepackage{hyperref} 
\usepackage{hyperxmp}
\usepackage{arxiv}
\usepackage[utf8]{inputenc} 
\usepackage[T1]{fontenc}    
\usepackage{hyperref}       
\usepackage{url}            
\usepackage{booktabs}       
\usepackage{amsfonts}       
\usepackage{nicefrac}       
\usepackage{microtype}      
\usepackage{lipsum}
\usepackage{graphicx}
\graphicspath{ {./images/} }
\usepackage{amsmath}
\usepackage{subcaption}
\usepackage{array}
\usepackage{tabularx}

\title{Beyond Charging Anxiety: An Explainable Approach to Understanding User Preferences of EV Charging Stations Using Review Data}

\author{
 Zifei Wang \\
  IMSE Department\\
  University of Michigan-Dearborn\\
  Dearborn, MI, 48128 \\
  \texttt{zifwang@umich.edu} \\
   \And
 Emmanuel Abolarin \\
  IMSE Department\\
  University of Michigan-Dearborn\\
  Dearborn, MI, 48128 \\
  \texttt{emmanue@umich.edu} \\
  \And
 Kai Wu \\
   Global Data Insight \& Analytics (GDI\&A)\\
   Ford Motor Company\\
   Dearborn, MI, 48124 \\
  \texttt{kwu41@ford.com} \\
  \And
Venkatarao Rebba \\
   Global Data Insight \& Analytics (GDI\&A)\\
   Ford Motor Company\\
   Dearborn, MI, 48124 \\
  \texttt{vrebba1@ford.com} \\
  \And
Jian Hu \\
  IMSE Department\\
  University of Michigan-Dearborn\\
  Dearborn, MI, 48128 \\
  \texttt{jianhu@umich.edu} \\
   \And
Zhen Hu \\
  IMSE Department\\
  University of Michigan-Dearborn\\
  Dearborn, MI, 48128 \\
  \texttt{zhennhu@umich.edu} \\
   \And
Shan Bao \\
  IMSE Department\\
  University of Michigan-Dearborn\\
  Dearborn, MI, 48128 \\
  \texttt{shanbao@umich.edu} \\
   \And
Feng Zhou \\
  IMSE Department\\
  University of Michigan-Dearborn\\
  Dearborn, MI, 48128 \\
  \texttt{fezhou@umich.edu} \\
}

\begin{document}
\maketitle

\begin{abstract}
Electric vehicles (EVs) charging infrastructure is directly related to the overall EV user experience and thus impacts the widespread adoption of EVs. Understanding key factors that affect EV users' charging experience is essential for building a robust and user-friendly EV charging infrastructure. This study leverages about $17,000$ charging station (CS) reviews on Google Maps to explore EV user preferences for charging stations, employing ChatGPT 4.0 for aspect-based sentiment analysis. We identify twelve key aspects influencing user satisfaction, ranging from accessibility and reliability to amenities and pricing. Two distinct preference models are developed: a micro-level model focused on individual user satisfaction and a macro-level model capturing collective sentiment towards specific charging stations. Both models utilize the LightGBM algorithm for user preference prediction, achieving strong performance compared to other machine learning approaches. To further elucidate the impact of each aspect on user ratings, we employ SHAP (SHapley Additive exPlanations), a game-theoretic approach for interpreting machine learning models. Our findings highlight the significant impact of positive sentiment towards "amenities and location", coupled with negative sentiment regarding "reliability and maintenance", on overall user satisfaction. These insights offer actionable guidance to charging station operators, policymakers, and EV manufacturers, empowering them to enhance user experience and foster wider EV adoption.
\end{abstract}



\section{Introduction}
The global transition towards sustainable transportation is rapidly accelerating, with electric vehicles (EVs) taking center stage, fueled by government incentives and the urgent need to mitigate greenhouse gas emissions \cite{USEPA}. A robust and user-friendly charging infrastructure is essential for the success of this transition, as it directly impacts the convenience and overall user experience, ultimately influencing EV adoption rates \cite{pevec2019electric}. Ensuring user satisfaction with charging stations is therefore paramount. 

While prior research has yielded valuable insights into aspects of EV charging, such as infrastructure optimization \cite{liu2018optimal, yi2022electric}, charging demand prediction \cite{zhao2017ev, shahriar2020machine}, and the influence of public charging availability on EV adoption \cite{chen2020assessing, haustein2021battery}, a crucial gap remains: a comprehensive understanding of the user experience based on real-world data, particularly one that moves beyond limited sample sizes in the current literature. Studies focusing on charging behavior and pricing strategies, like those by Morrissey and Weldon \cite{morrissey2016future} and Jiang et al. \cite{jiang2021review}, primarily rely on objective data or simulated scenarios, often overlooking the nuanced sentiments and preferences of EV users. These approaches, while informative, may not accurately reflect the diversity and complexity of real-world EV charging experiences. Furthermore, reliance on survey data or qualitative studies with limited sample sizes, as often seen in studies exploring user perceptions \cite{pevec2020survey, globisch2019consumer}, hinders the generalizability of findings and limits our ability to draw robust conclusions about user preferences across broader populations. As highlighted by Sun et al. \cite{sun2016fast}, there is a surprising dearth of research dedicated to exploring user perceptions of and preferences for charging in the field of EV acceptance, especially using large-scale, real-world data. This gap is especially significant considering that charging represents a fundamental departure from the familiar experience of fueling a gasoline vehicle, potentially influencing user satisfaction and adoption decisions. 

This study addresses this critical gap by introducing an innovative approach that leverages the power of large-scale review data and advanced machine-learning techniques. We analyze a comprehensive dataset of charging station reviews on Google Maps, harnessing the rich and authentic user-generated content reflecting real-world EV charging experiences. To extract fine-grained sentiment towards specific aspects of charging stations, we employed ChatGPT 4.0, a state-of-the-art large language model, to conduct aspect-based sentiment analysis (ABSA). This methodology allowed us to uncover the nuances of user preferences, revealing key points that traditional methods might miss. 

Furthermore, we developed a robust and explainable preference model using LightGBM \cite{ke2017lightgbm, zhou2021using}, a highly efficient gradient boosting framework. This model was able to predict user ratings (i.e., preferences) based on the identified aspects and their associated polarities based on aspect-based sentiment analysis using ChatGPT, enabling us to understand user satisfaction at both individual and aggregated levels. To ensure model interpretability, we utilized SHAP (SHapley Additive exPlanations) \cite{lundberg2020local, ayoub2022predicting}, providing a clear understanding of the factors driving user preferences. By combining large-scale review data with advanced machine learning techniques, this study provided actionable insights for stakeholders in the EV charging infrastructure industry, empowering them to design a more user-centric charging experience and ultimately accelerate the transition towards sustainable transportation.

In summary, our study research makes the following key contributions to the field of EV charging station research:
\begin{itemize}
    \item Large-Scale Real-World Data Analysis: Overcomes limitations of previous studies by analyzing a vast dataset of Google Maps reviews, capturing diverse user experiences across a wider population.
    \item Nuanced Sentiment Extraction: Employs ChatGPT 4.0 for aspect-based sentiment analysis (ABSA), revealing fine-grained user sentiment towards specific aspects of charging stations. 
    \item Explainable Preference Prediction: Develops a robust and explainable LightGBM-based preference model for predicting user preferences, enhanced with SHAP analysis for interpretable insights into driving factors.
\end{itemize}

The remainder of this paper is organized as follows: Section 2 provides an overview of related work, highlighting key studies and existing knowledge relevant to the research. Section 3 details the methodology, including data collection, sentiment analysis, and preference modeling. Section 4 presents the results obtained from the analyses. Section 5 discusses the implications of the findings, addressing limitations and potential future directions. Finally, Section 5 concludes the paper. 

\section{Related Work}
\subsection{Factors Influencing Charging Behavior}
Numerous studies have investigated the drivers behind EV charging behavior, revealing a complex interplay of individual preferences, household attributes, and external factors. Demographic factors, such as age, income, and education level, have been shown to play a significant role in EV adoption and subsequently influence charging patterns. For instance, Nazari et al. \cite{nazari2023electric} found that age significantly influenced EV adoption and transaction decisions, while Bruckmann et al. \cite{BRUCKMANN2023103626} revealed that consumer preferences towards public charging infrastructure varied with age and education level. Vehicle attributes, including range and charging speed, also impact charging decisions, with users of long-range EVs with fast charging capabilities exhibiting different behaviors compared to those with limited range and slow charging options \cite{Nicholas_et_al_2017, Bansal_et_al_2021, Danielis_et_al_2020, Jensen_et_al_2021}. Social networks and environmental consciousness have been identified as additional determinants, with environmentally conscious individuals demonstrating a greater inclination toward sustainable charging practices. For example, Jensen et al. \cite{Jensen_et_al_2014} found that EV adopters exhibited a stronger affinity for green energy solutions and prioritized using sustainable energy sources for charging.

Beyond individual preferences, researchers emphasized the impact of daily mobility patterns, economic factors, and geographical context on charging behavior. Variables such as dwell time at destinations, vehicle miles traveled, and time of day significantly influence charging frequency and location choices. John et al. \cite{John20222147041} highlighted the influence of daily mobility patterns and work schedules on charging behavior among German EV users. Meanwhile, Franke and Krems \cite{Franke_and_Krems_2013} explored how range anxiety, convenience, and charger location awareness impacted daily charging decisions. Regional infrastructure disparities, policy incentives, and electricity pricing structures further contributed to the complexity of charging behavior, highlighting the need to consider geographical context. For instance, Egnér and Trosvik \cite{Egnér_and_Trosvik_2018} demonstrated the impact of local policy instruments and regional charging infrastructure availability on EV adoption rates in Sweden.

Recognizing the crucial role of accessible and strategically located charging infrastructure, researchers also focused on optimizing site selection strategies. Previous studies investigated the influence of road types, urban accessibility, parking availability, and proximity to traditional fuel stations on charging station placement \cite{mahdy2022multi, KLOS2023103601, CHARLY2023104573}. Availability of amenities, such as restrooms, restaurants, and shopping facilities, were shown to increase the attractiveness of charging locations, influencing user choices. This was highlighted by Figenbaum and Nordbakke \cite{Figenbaum_and_Nordbakke_2019}, who found that Norwegian EV users considered amenities and surrounding facilities as key factors when selecting charging locations. The multifaceted nature of site selection necessitates a comprehensive evaluation of both direct and indirect costs, encompassing factors such as land value, accessibility, grid capacity, and potential user revenue, to ensure sustainable and profitable charging infrastructure. Li et al. \cite{li2022improved} proposed an improved whale optimization algorithm for location selection, emphasizing the importance of considering both direct infrastructure costs and indirect factors like land value and user revenue.

While these studies provide valuable insights into the objective factors influencing charging behavior and infrastructure placement, they often relied on traditional survey methods or qualitative approaches with limited sample sizes, hindering the generalizability of their findings. Furthermore, a critical gap persists in understanding the subjective user experience and the nuanced preferences that influence charging station choices.

\subsection{Predictive Modeling of Charging Behavior}
The advent of machine learning has revolutionized our ability to understand and predict EV charging behavior. Researchers have leveraged powerful algorithms, including decision trees, random forests, support vector machines, and neural networks, to model charging demand based on various factors such as vehicle characteristics, user demographics, charging infrastructure availability, and temporal patterns \cite{liu2019brief, liu2019public, jiang2021review}. Ensemble methods like random forests have demonstrated effectiveness in capturing nonlinear relationships and uncertainties inherent in charging behavior, leading to more accurate predictions \cite{ibrahim2020machine}. Furthermore, integrating contextual information, such as weather conditions, electricity prices, and traffic patterns, can further enhance the predictive capabilities of these models \cite{zhao2017ev}. Clustering and classification algorithms offer a means to segment users based on their charging patterns and preferences, enabling tailored recommendations and incentives to promote efficient and sustainable charging practices \cite{xu2018planning, ferguson2018optimal, cao2021smart}. However, despite the predictive power of these machine learning models, they often lack the interpretability and explainability required to understand the underlying factors driving user preferences.

In our study, we addressed the research gaps by leveraging a large-scale dataset of Google Maps reviews and employing cutting-edge machine learning techniques, coupled with explainability tools, to uncover the nuanced factors driving user satisfaction with EV charging stations. Our approach moves beyond objective factors and traditional methodologies, offering a deeper understanding of the subjective user experience and paving the way for a more user-centric and efficient EV charging infrastructure.


\section{Methods}
This study aims to address the critical gap in understanding user preferences towards EV charging stations by leveraging a large-scale dataset of Google Maps reviews and employing cutting-edge machine learning techniques coupled with explainability tools. Our research objectives are to: (1) identify key aspects influencing user satisfaction with EV charging stations, (2) develop a robust and accurate preference model for predicting user ratings, and (3) provide interpretable insights into the factors driving those preferences. To achieve these objectives, we employ a systematic and data-driven approach involving the following steps:

\subsection{Data Collection}

We collected a comprehensive dataset of 16,768 public available user reviews of charging stations from Google Maps, using the Google Maps API. These reviews contain valuable information, including the names and addresses of charging stations, user ratings (on a scale of 1 to 5), and the review texts, providing rich insights into user experiences and preferences.

\subsection{Data Validation and Quality Assurance}
To address concerns regarding the validity, reliability, and potential biases of user-generated content from Google reviews, we implemented a rigorous data validation and quality assurance protocol during preprocessing. Each raw review record includes a unique user identifier (author\_url), a precise timestamp, the review text, and the geographic coordinates (latitude and longitude) of the corresponding charging station.
First, to mitigate the risk of data duplication or automated entries, we enforced uniqueness at the granular level. Only reviews representing a distinct combination of user identifier, timestamp, and charging station location were retained. This ensures that individual reviews reflect unique interactions and prevents overrepresentation from redundant entries by the same user.

Second, we analyzed user contribution patterns to assess potential concentration of activity or bot-like behavior. Out of 12,601 valid reviews, 9,609 distinct users were identified. A substantial majority, 8,061 users (approximately 83.9\%), contributed only a single review, indicating a broad and diverse base of contributors. Users with multiple contributions were scarce: 959 posted twice, 310 posted three times, and 107 posted four times. The most active user submitted 42 reviews across various locations and timepoints; these were manually examined and confirmed to be distinct, contextually meaningful, and reflective of genuine diverse experiences with the EV charging network.

Third, we performed an in-depth analysis of textual uniqueness to identify potential automated or low-information content. Out of 12,601 valid reviews, 12,553 (99.6\%) contained unique text content. Only seven short phrases appeared more than three times (e.g., 'Great location' appeared 8 times, 'Out of order' appeared 5 times), and 27 review texts appeared exactly twice. This means 12,519 reviews (99.3\%) were completely unique in their textual content, strongly supporting the diversity and richness of the dataset and significantly reducing concerns regarding content automation or repetition.

While direct ground truth validation of individual user identities is not feasible due to privacy constraints inherent in Google Maps data, our multi-pronged approach — combining stringent filtering for uniqueness, detailed analysis of user contribution frequencies, and comprehensive examination of textual diversity — provides strong empirical assurances regarding the quality and representativeness of our review dataset.

Finally, in terms of temporal scope, our dataset predominantly comprises recent user experiences, with approximately 87\% of all reviews posted between 2021 and 2023. Specifically, 5,412 reviews are from 2022, 3,536 from 2023, and 1,700 from 2021. Reviews from earlier years contribute less significantly, ensuring that our analysis reflects contemporary user interactions with the evolving EV charging infrastructure. Although explicit demographic information is not available, the inclusion of latitude and longitude for each charging station allows for an implicit understanding of geographic context, aiding in the interpretation of results within specific regional charging environments.

\subsection{Aspect-Based Sentiment Analysis}

To extract meaningful insights from the unstructured review texts, we employed aspect-based sentiment analysis (ABSA) using ChatGPT 4.0, a state-of-the-art large language model (LLM) capable of understanding and generating human-like text \cite{zhou2020machine}. Recent studies have demonstrated the effectiveness of LLMs in a variety of information extraction tasks, particularly in low-resource or domain-specific contexts where annotated data are limited \cite{zhang2024exploring, xu2024exploring}. Compared to traditional rule-based or supervised learning methods, LLMs offer greater flexibility, support for multi-aspect extraction, and robustness to informal and noisy language, which are especially common for real-world user-generated reviews \cite{huang2024critical}. This process involved several steps:

\textbf{Initial AI Analysis:} We conducted an initial analysis using ChatGPT 4.0 with carefully tuned prompts to identify primary aspects and their sentiments within the review data.

\textbf{Manual Examination and Consolidation:} We manually examined and consolidated the identified aspects into 12 major categories relevant to EV charging stations (as detailed in Table \ref{tab:Definitions}). 

\textbf{Ground Truth Labeling:} A subset (1,890) of the user reviews was manually coded to establish a ground truth for model evaluation and fine-tuning.

\textbf{AI Refinement with Few-Shot Learning:} Leveraging the insights from the manual coding, we fine-tuned the prompts provided to ChatGPT 4.0, incorporating few-shot examples to enhance its ability to accurately identify aspects and sentiments.

\textbf{Model Evaluation and Application:} We evaluated the performance of the fine-tuned ChatGPT model against the manually coded ground truth to ensure the credibility of the automated text analysis. The refined model was then applied to predict aspects and sentiments in the remaining corpus of reviews.

To further clarify the AI refinement with few-shot learning step, we designed a task-specific prompt to guide ChatGPT-4.0 in extracting aspect-sentiment pairs from each user review with the following features:
\begin{itemize}
    \item \textbf{Standardized instruction-based prompt:} We developed a precise, instruction-based prompt to guide the LLM's behavior and ensure uniformity in processing diverse review texts.
    \item \textbf{Aspect constraints:} The model is explicitly instructed to categorize review content only within a predefined set of 12 relevant EV charging station aspects, preventing the generation of extraneous or irrelevant categories.
    \item \textbf{Fixed output format:} We mandated a consistent output format of [aspect/polarity/summary] for each extracted sentiment, which facilitates subsequent quantitative and qualitative analysis.
    \item \textbf{Few-shot learning examples:} The prompt incorporates a few-shot learning approach by providing 20 carefully curated labeled examples. These examples are critical as they guide the LLM in both the task structure and the nuanced interpretation required for accurate sentiment and aspect identification, particularly for multi-topic reviews.
\end{itemize}

\begin{figure*}
     \centering
     \includegraphics[width=0.8\linewidth]{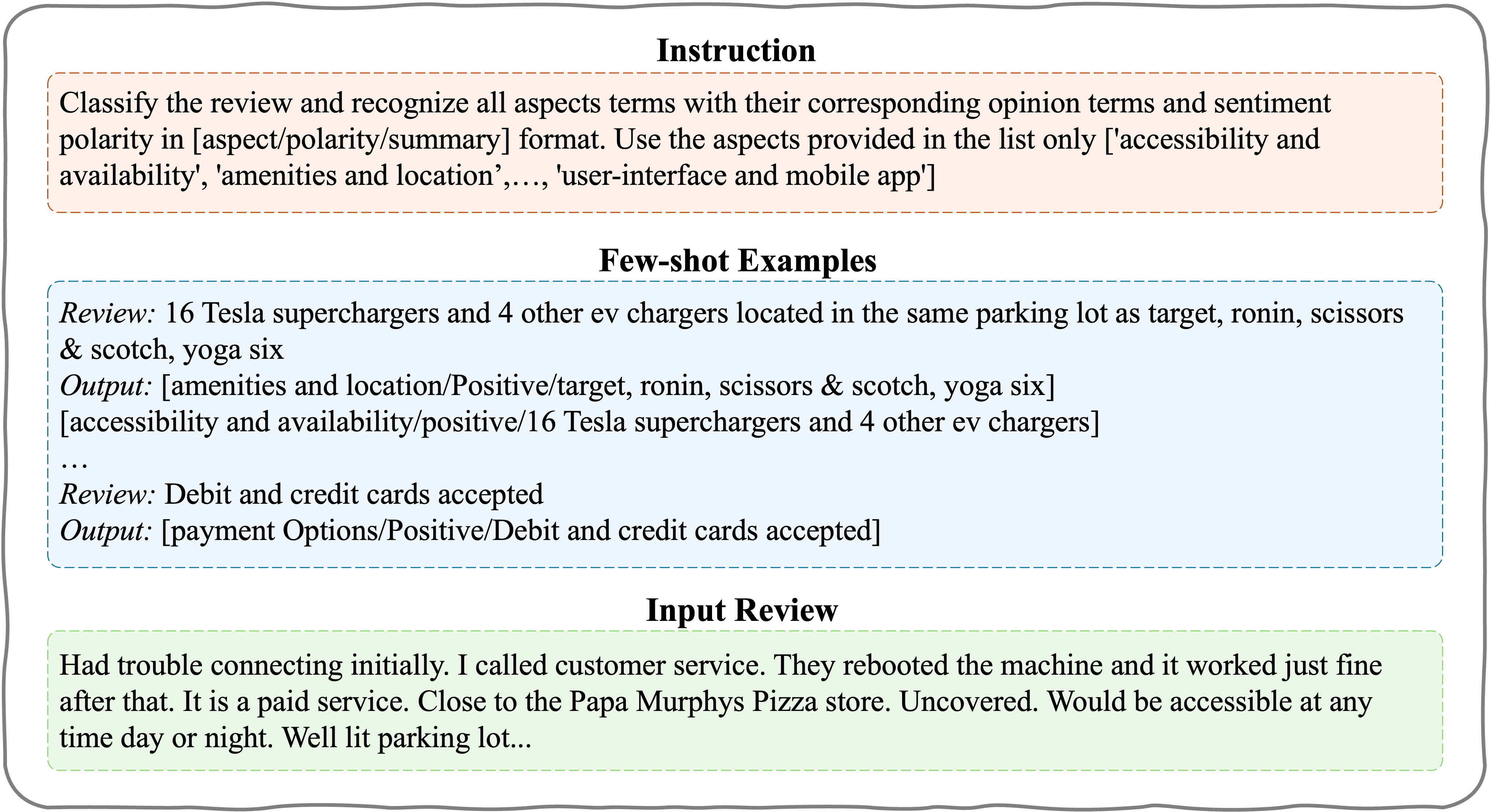}
     \caption{Example of the Structured Prompt used for ABSA}
     \label{fig:prompt}
\centering
\end{figure*}

A representative example that shows the full prompt and the corresponding examples is provided in Fig.~\ref{fig:prompt}. This structured prompting strategy enabled fine-grained, interpretable extraction of multiple aspect-level sentiments from unstructured user reviews.

\begin{table*}[t]
\centering{
\caption{Key Aspects Influencing User Satisfaction with EV Charging Stations}
\label{tab:Definitions}
\hspace*{-1.6cm} 
\begin{tabular}{c l m{6.5cm} m{6.3cm}}
\toprule
\multicolumn{1}{c}{\textbf{No.}} & 
\multicolumn{1}{c}{\textbf{Aspect}} & 
\multicolumn{1}{c}{\textbf{Definition}} & 
\multicolumn{1}{c}{\textbf{Example}} \\
\midrule
 1& Accessibility and Availability & The availability and ease of reaching and utilizing the charging infrastructure for electric vehicles. &  ``Not accessible to the public", ``Only available for guests", ``Plenty of available charging stalls" \\ 
 2& User-interface and Mobile App & The graphical or physical means through which users interact and control the charging process. & ``Easiest interface", ``Confusing and not user-friendly" \\   
 3& Queue and Waiting Time & The time a driver must wait for an available charging spot. &  ``No wait time", ``Longer waiting times on weekends", ``Limited availability" \\  
 4& Amenities and Location & Additional facilities or services provided at the charging location. & ``Access to restrooms and snacks", ``Easy to walk to nearby restaurants" \\  
 5& Charging Speed and Efficiency & The rate at which the charger replenishes the battery's energy. & ``Takes 40 minutes to fully charge", ``Disappointed with slow charging speed" \\ 
 6& Reliability and Maintenance & The consistent performance and upkeep of the charging infrastructure. & ``Chargers not functioning", ``Well-maintained charging station", ``Chargers often down" \\ 
 7& Compatibility and Connectivity & The ability of the charger to effectively communicate and deliver power to the EV battery. &  ``Prius compatible, fast charge", ``Tesla model 3 issue", ``Cable doesn't reach" \\  
 8& Safety & Measures taken to prevent risks and hazards associated with the charging process. & ``Unsafe behavior observed", ``Secure with a guard and security gate" \\  
 9& Payment Options & The methods or systems available for users to pay for charging. &  ``Pay station on ground floor doesn't work", ``Debit and credit cards accepted" \\  
 10& Ease of Use & The level of convenience and user-friendliness provided by the charging station. & ``Difficult to initiate", ``Simple to use", ``Poorly designed charging stations" \\   
 11& Price and Cost & The monetary cost of charging, including any additional fees. & ``Expensive overnight charge", ``Charged to 100\% in two hours for \$20", ``Inefficient charging" \\  
 12& Customer Service & The quality of assistance and support provided to EV owners. & ``Exceptional service and driver support", ``Helpful customer support" \\  
 \bottomrule
\end{tabular}
}
\end{table*}

\subsection{Preference Modeling with Explainability}

To develop a robust and accurate model for predicting user ratings and understanding the key factors influencing those ratings, we constructed two distinct preference models: a micro-level model focusing on individual user satisfaction and a macro-level model capturing collective sentiment towards specific charging stations.

\subsubsection{Data Processing}

The output from the aspect-based sentiment analysis, consisting of 34,211 aspects with polarity across 16,768 reviews, among which were 13,618 distinct review texts and 5,268 distinct charging stations, served as the foundation for our preference models. To create structured datasets suitable for modeling, we processed the data into two distinct formats:

\textbf{Micro-Level Dataset}: This dataset focused on individual user ratings ($n = 13,618$). For each review, we created a binary representation of the 12 identified aspects, with separate columns for positive and negative polarities. If an aspect with a specific polarity was mentioned in the review, the corresponding column in the dataset was assigned a value of 1. Conversely, if the aspect-polarity combination was not identified in the review, the corresponding column received a value of 0. This resulted in a total of 24 columns, capturing the presence or absence of each aspect-polarity combination within the review text (see Table~\ref{tab: individual_review} for an example).  


\begin{table*}[htbp]
\centering
\caption{Example of Micro-level Individual User Review Structured Data}
\label{tab: individual_review}
\hspace*{-0.7cm}
\begin{tabular}{c c c c m{3.3cm} m{3.3cm} c}

\toprule
\textbf{Charging Station} & \textbf{Address} & \textbf{Rating} & \textbf{Review} & \textbf{Accessibility and availability\_Negative} & \textbf{Accessibility and availability\_Positive} & \textbf{...} \\ 
\midrule
7-Eleven & 5745 Clark... & 5 & Downloaded... 
& 0 & 0 & ... \\

7Charge & 5301 34th... & 4 & Was ICE'd...
& 1 & 0 & ...\\

AmpUp & 1 Union... & 3 & Inside the...
& 0 & 0 & ...\\

Applegreen Electric & 1 South... & 1 & Could not use... 
& 1 & 0 & ... \\

Blink & 1027 Putney... & 3 & Single car... 
& 0 & 1 & ... \\

\bottomrule
\end{tabular}
\end{table*}

\textbf{Macro-Level Dataset}: This dataset captured the collective sentiment towards each charging station ($n = 5,268$).  We aggregated the individual reviews associated with each station and computed the mean value for the 24 aspect-polarity columns, along with the average user rating for each specific charging station. This provided a comprehensive view of the perceived strengths and weaknesses of each charging station 
(see Table~\ref{tab: averageCS_review} for an example). In addition to the average rating and aspect-based sentiment features, two supplementary columns are included in Table~\ref{tab: averageCS_review} to enhance interpretability: Rating Count, which represents the number of individual user reviews aggregated for each charging station, and Rating Standard Deviation, which reflects the variability of ratings at that station.

\begin{table*}[h]
\centering{
\caption{Example of Macro-level Averaged Charging Station Structured Data}
\label{tab: averageCS_review}
\hspace*{-0.9cm} 
\begin{tabular}{c m{2cm} m{1cm} m{1cm} m{1cm} m{3.5cm} m{3.5cm} c}

\toprule
\textbf{Charging Station} & \textbf{Address} & \textbf{Rating Avg.} & \textbf{Rating Count} & \textbf{Rating SD} & \textbf{Accessibility and availability\_Negative} & \textbf{Accessibility and availability\_Positive} & \textbf{...} \\ 
\midrule
Blink & 2000 Kille... & 2.0 & 5 & 1.4 
& 0.2 & 0.0 & ... \\

ChargePoint & 101 Stego... & 2.5 & 4 & 1.7 
& 0.0 & 0.0 & ...\\

Big Sky & 25 Town... & 5.0 & 2 & 0.0 
& 0.0 & 1.0 & ...\\

EVgo & 5900 Suga... & 2.3 & 3 & 1.5 
& 0.0 & 0.3 & ... \\

Electrify & 45550 Dulle... & 4.5 & 4 & 1.0 
& 0.0 & 0.3 & ... \\

\bottomrule
\end{tabular}
}
\end{table*}

These two complementary datasets provided a multi-faceted perspective on user sentiments and preferences, enabling us to develop a more comprehensive understanding of the factors influencing user satisfaction with EV charging stations.

\subsubsection{Predicting User Preference Using LightGBM}

To predict user ratings based on the identified aspects and their associated sentiments, we employed LightGBM (Light Gradient Boosting Machine), a high-performance implementation of the gradient boosting framework \cite{ke2017lightgbm}. Gradient boosting is a general ensemble learning technique, which builds predictive models in a sequential manner \cite{natekin2013gradient}. At each iteration, a new decision tree is trained to predict the residual errors (i.e., the difference between the actual user rating and the current model's prediction in our study), which means the part the current model has not yet captured \cite{ahn2023ensemble}. LightGBM implements this approach and constructs an ensemble of decision trees iteratively, with each new tree added to minimize the residuals between the actual target values (user ratings) and the predictions of the current ensemble. This iterative optimization process gradually improves the model's predictive performance. Compared to other methodologies, LightGBM is particularly renowned for its efficiency, accuracy, and scalability \cite{shehadeh2021machine}.

The model's prediction at the $t_{th}$ iteration for the $i_{th}$ instance, denoted as $\hat{y}_i^\text{(t)}$, is calculated by adding the residual of the corresponding tree $f_t(\cdot)$ to the prediction from the previous iteration:

\begin{equation}
    \hat{y}_i^\text{(t)} = \hat{y}_i^\text{(t-1)} + f_{t}({\bf{x}}_{i}),
\end{equation}

\noindent where ${\bf{x}}_i = (x_{i1}, x_{i2}, ..., x_{ip})$ represents the input vector of $p$ predictor variables (in our case, the 24 aspect-polarity combinations).  

The model is trained by minimizing the following objective function:

\begin{equation}
    \label{eq:loss}
    L^\text{(t)} = \sum_{i}^{N} {l(y_i,\hat{y}_i^\text{(t)})} + \sum_{t=1}^{T} {\Omega(f_t)},
\end{equation}

\noindent where $l$ is the loss function measuring the difference between the actual rating $y_i$ and the predicted rating $\hat{y}_i^\text{(t)}$, and $\Omega(f_t)$ is a penalty term applied to each node to control the complexity of the trees and to reduce the risk of overfitting. $N$ denotes the total number of data samples in the dataset, while $T$ indicates the total number of trees generated by a LightGBM model.

To further enhance performance and handle large-scale data, LightGBM incorporates two novel techniques: Gradient-based One-Side Sampling (GOSS) and Exclusive Feature Bundling (EFB) \cite{ke2017lightgbm, wen2021quantifying}. GOSS improves training speed and reduces memory usage by strategically sampling instances based on their gradient values, while EFB increases feature selection efficiency by bundling mutually exclusive features. These techniques enable LightGBM to effectively handle high-dimensional data and achieve superior predictive accuracy.

\subsubsection{Explaining User Preference Using SHAP}

To gain interpretable insights into the factors driving user preferences, we utilized SHAP (SHapley Additive exPlanations), a powerful method rooted in game theory that explains the output of machine learning models by attributing the contribution of each feature to the model's predictions \cite{lundberg2020local}. It uses the concept of Shapley values, which was originally used to determine the fair contributions among players, to machine learning models by treating input features as players and the model prediction as the result \cite{shapley1953value, molnar2020interpretable}. SHAP values quantify the marginal contribution of each feature to the difference between the model's prediction for a specific instance and the average model prediction.

SHAP achieves this by constructing a linear explanation model, denoted as $g(x')$, which approximates the original machine learning model, $f(x)$:
\begin{equation}
    g(x') = f(x) = \phi_{0} + \sum_{m=1}^{M} {\phi_{m}x'_{m}},
\end{equation}
\noindent where $g(x')$ is the explanation model prediction for simplified input $x'$; $f(x)$ is the model prediction for input $x$; $M$ is the number of simplified input features; $\phi_{0}$ is the expected value of original model's output, and $\phi_{m}$ represents the effect of the $m_{th}$ feature on the model prediction.

Three unique properties are desired by the solution of the additive feature attribution methods, which are local accuracy, missingness, and consistency. Firstly, the local accuracy is achieved by matching the original model \textit{f(x)} and the explanation model \textit{g(x')}, when the simplified inputs \textit{x'} is mapped to the original inputs \textit{x} through the mapping function \(x=h_{\text{x}}(x')\). Then the missingness ensures that when the simplified input indicates the absence of that feature (i.e., \(x_{\text{n}}'=0\)), then the impact of that feature should be zero (i.e., \(\phi_{\text{n}}=0\)). Lastly, the third property, consistency, indicates that the feature input's attribution should not decrease if the changes in the simplified input increase or persist in the model results. The three desired properties could only be satisfied by only one explanation model $\textit{g}(\cdot)$, where the Shapley values originated from the combined cooperative game theory are expressed as:
\begin{equation}
    \label{eq:Shapley value}
    \phi_{n}(f, x) = \sum_{z' \subseteq x'} {\frac{|z'|!(N-|z'|-1)!}{N!}[f_{x}(z')-f_{x}(z'\backslash n)] },
\end{equation}
\noindent where $|z'|$ is the number of non-zero features in $z$.

\section{Results}

\subsection{Identifying Key Aspects of EV Charging Stations}

\begin{figure*}
     \centering
     \includegraphics[width=0.7\linewidth]{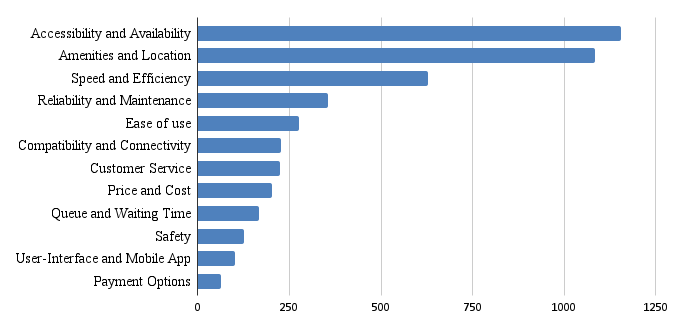}
     \caption{Twelve Key Aspects of EV Charging Station Experience}
     \label{fig:12_Aspects}
\centering
\label{fig:expProc}
\end{figure*}

Our initial analysis using ChatGPT 4.0 identified 112 distinct aspects or topics within the 5750 manually coded user reviews. These aspects varied significantly in their frequency of occurrence, with some emerging prominently across the dataset while others were mentioned less frequently. To create a more manageable and interpretable set of aspects, we manually examined and consolidated these 112 aspects into 12 major categories, as summarized in Table~\ref{tab:Definitions} and illustrated in Fig.~\ref{fig:12_Aspects}.

These 12 aspects encompass a wide range of factors that are likely to influence user satisfaction with EV charging stations, providing a comprehensive framework for analyzing user sentiment and preferences.

\subsection{ABSA Model Performance Evaluation}

To establish a reliable benchmark for evaluating the performance of our ABSA model, we created a set of ground truth labels. We selected 250 overlapping reviews from the 5,750 manually coded reviews and had three trained experts independently annotate them. This process resulted in an initial agreement rate of 0.74, subsequent researcher audits and discussions among the experts resolved discrepancies, leading to a reliable set of ground truth labels.

To validate the accuracy and reliability of our ABSA using ChatGPT 4.0, we separately compared its performance against the manually coded ground truth labels for a subset of 1,857 and 1,890 reviews for aspects and polarity respectively. The model achieved a relatively high performance across various evaluation metrics, including precision, recall and f1-score: aspect (0.85, 0.89, 0.85) and polarity (0.87, 0.87, 0.87) (see Tables \ref{tab: performance evaluation 1} and \ref{tab: performance evaluation 2}). These results demonstrate the effectiveness of our approach in accurately identifying and categorizing user sentiments towards specific aspects of EV charging stations.

\begin{table*}[t]
\centering
\caption{Performance of ChatGPT 4.0 for Aspect-Based Sentiment Analysis}
\label{tab: performance evaluation 1}
\begin{tabular}{lcccc}
\toprule
\textbf{Aspect} & \textbf{Precision} & \textbf{Recall} & \textbf{F1-score} & \textbf{Support} \\ \midrule
Accessibility and Availability & 0.84 & 0.79 & 0.81 & 410 \\
User-interface and Mobile App & 0.86 & 1.00 & 0.92 & 12 \\
Queue and Waiting Time & 0.82 & 0.97 & 0.89 & 97 \\
Amenities and Location & 0.86 & 0.89 & 0.89 & 424 \\
Charging Speed and Efficiency & 0.83 & 0.82 & 0.82 & 311 \\
Reliability and Maintenance & 0.78 & 0.91 & 0.84 & 254 \\
Compatibility and Connectivity & 0.76 & 0.98 & 0.86 & 109 \\
Safety & 0.75 & 0.86 & 0.80 & 49 \\
Payment Options & 0.95 & 1.00 & 0.97 & 18 \\
Ease of Use & 0.78 & 0.96 & 0.86 & 26 \\
Price and Cost & 0.81 & 0.83 & 0.82 & 89 \\
Customer Service & 0.78 & 0.98 & 0.87 & 91 \\
\hline
Micro Average & 0.82 & 0.87 & 0.84 & 1890 \\
Macro Average & 0.82 & 0.92 & 0.86 & 1890 \\
Weighted Average & 0.82 & 0.87 & 0.84 & 1890 \\
\bottomrule
\end{tabular}
\end{table*}

\begin{table}[h]
\centering
\caption{Performance of ChatGPT 4.0 for Polarity-Based Sentiment Analysis}
\label{tab: performance evaluation 2}
\begin{tabular}{lcccc}
\toprule
\textbf{Aspect} & \textbf{Precision} & \textbf{Recall} & \textbf{F1-score} & \textbf{Support} \\ \midrule
Positive & 0.87 & 0.93 & 0.90 & 1075 \\
Negative & 0.93 & 0.79 & 0.85 & 823 \\
\hline
Micro Average & 0.89 & 0.87 & 0.88 & 1898 \\
Macro Average & 0.90 & 0.86 & 0.87 & 1898 \\
Weighted Average & 0.89 & 0.87 & 0.88 & 1898 \\
\bottomrule
\end{tabular}
\end{table}

\subsection{Predicting User Satisfaction with LightGBM}

The ChatGPT model generated 34,211 aspects with polarity after removing the invalid outputs (e.g., aspects not identified or polarity not identified) among 12,601 individual user reviews and 4,516 charging stations from the raw review dataset. These voluminous results allowed for a thorough exploration of diverse customer sentiments and preferences regarding EV charging stations. The distribution of the aspects with polarity was shown in Fig.~\ref{fig:aspect_count}.

It is important to note that the input features for the LightGBM model, representing the presence or absence of specific aspect-polarity pairs (e.g., 'Accessibility and Availability/Positive', 'Charging Speed and Efficiency/Negative'), exhibited an imbalanced distribution. As depicted in Figs.~\ref{fig:12_Aspects} and ~\ref{fig:aspect_count}, certain aspects, such as 'Accessibility and Availability' and 'Amenities and Location,' appeared substantially more frequently in user reviews than less commonly mentioned aspects like 'Payment Options' or 'Safety.' This imbalance is not treated as a deficiency, but rather as a direct representation of the real-world emphasis and priorities within user feedback on EV charging stations. Preserving this natural distribution is crucial for maintaining the authenticity and interpretability of user concern patterns.

Given our primary objective of interpreting the influence of each aspect on user satisfaction, rather than solely maximizing predictive accuracy, we specifically chose not to artificially balance the feature set through techniques like resampling or reweighting. Such interventions, while useful for improving classification performance on minority classes, could distort the genuine prevalence and impact of various aspects in real-world user experiences. Instead, we relied on SHapley Additive exPlanations (SHAP) for model interpretation. SHAP's theoretical framework provides a robust and fair method for attributing the contribution of each feature to the model's output, enabling reliable assessment of feature importance even with imbalanced input features. This allows us to accurately discern which aspects, regardless of their frequency, significantly influence the predicted user satisfaction ratings, thereby providing authentic insights into user preferences.

 \begin{figure*}
     \centering
     \includegraphics[width=0.8\linewidth]{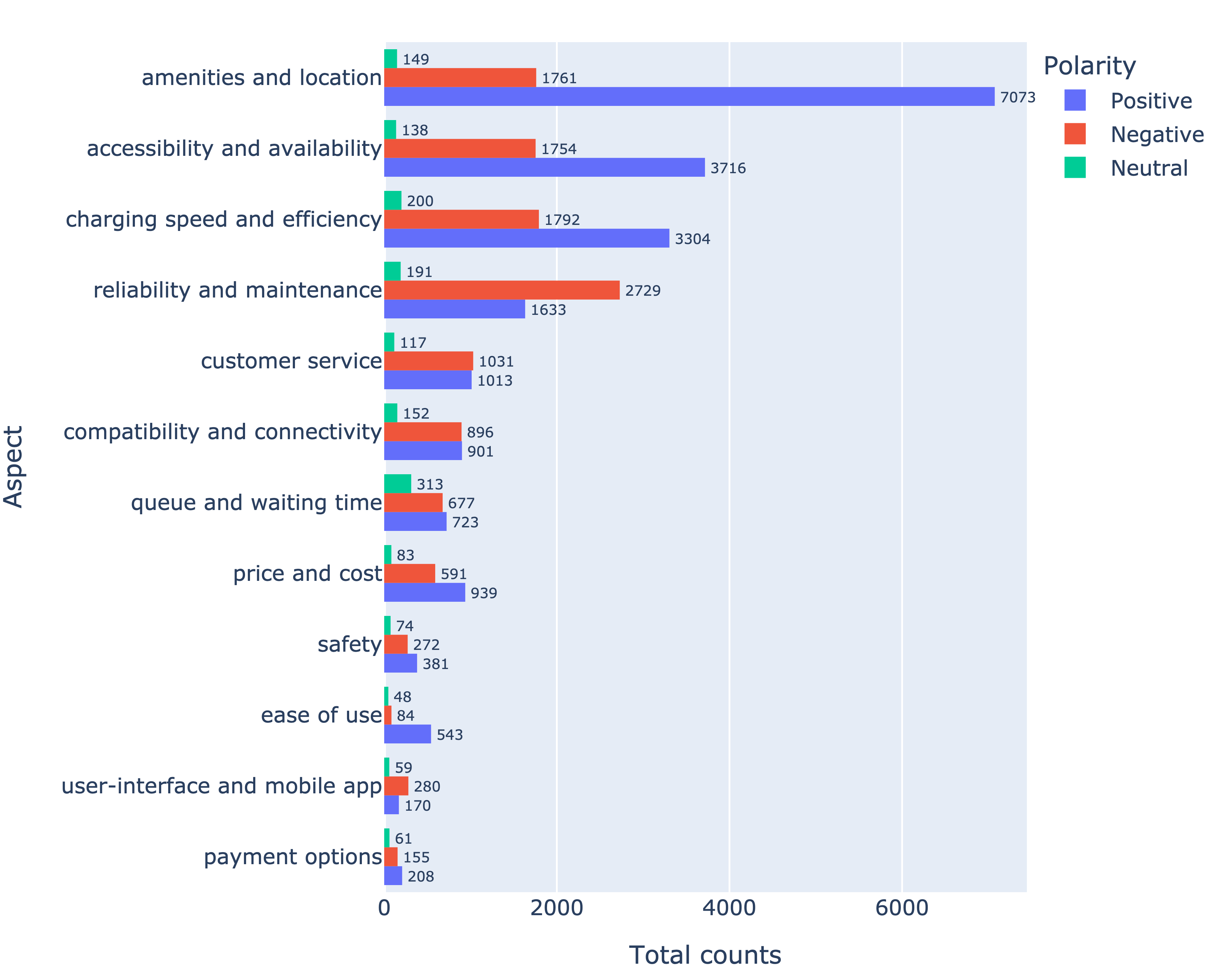}
     \caption{Distribution of Aspect with Polarity from the Google Review Dataset}
     \label{fig:aspect_count}
\centering
\label{fig:aspect_count}
\end{figure*}

We applied LightGBM, a powerful gradient boosting algorithm, to the processed data to predict user ratings based on the identified aspects and their polarities. We evaluated the model's performance using standard metrics, including Mean Absolute Error (MAE), Root Mean Squared Error (RMSE), and the correlation coefficient between predicted and actual ratings:

\textbf{Mean Absolute Error (MAE):}
\begin{equation}
\text{MAE} = \frac{1}{n} \sum_{i=1}^{n} |y_i - \hat{y}_i|
\end{equation}

\textbf{Root Mean Squared Error (RMSE):}
\begin{equation}
\text{RMSE} = \sqrt{\frac{1}{n} \sum_{i=1}^{n} (y_i - \hat{y}_i)^2}
\end{equation}

\textbf{Correlation Coefficient:}
\begin{equation}
r = \frac{\sum_{i=1}^{n} (y_i - \bar{y})(\hat{y}_i - \bar{\hat{y}})}{\sqrt{\sum_{i=1}^{n} (y_i - \bar{y})^2 \sum_{i=1}^{n} (\hat{y}_i - \bar{\hat{y}})^2}}
\end{equation}

where \( n \) equals the total number of data samples; \( y_i \) represents the actual value for the \( i \)th rating data sample; \( \hat{y}_i \) represents the predicted value for the \( i \)th data sample; \( \bar{y} \) is the mean of the actual values and \( \bar{\hat{y}} \) is the mean of the predicted values.

\subsubsection{Micro-Level Model Performance}

The micro-level model focused on individual user ratings. A 10-fold cross-validation was applied to validate the model results. A selected set of regression models, including linear regression, linear Support Vector Machine (SVM), Decision Tree, Random Forest, and XGBoost, were compared with the LightGBM models with the two predictor variable configurations. 
The performance comparison of the models was illustrated in Table~\ref{tab:Table 2}, achieving a MAE of 0.590, an RMSE of 0.857, and a correlation coefficient of 0.912. It is noted that the LightGBM is the most effective and accurate model, with the best model performance. These results demonstrate the model's high accuracy in predicting individual user satisfaction based on the sentiments expressed towards various aspects of charging stations.


\begin{table*}[h] 
\centering
\caption{Performance of selected regression models with 24 aspect with polarity predictor variables.}
\begin{tabular}{ccc|cc|cc}
\toprule
Model & \multicolumn{2}{c|}{MAE} & \multicolumn{2}{c|}{RMSE} & \multicolumn{2}{c}{Corr.} \\
& Mean & SD & Mean & SD & Mean & SD \\
\midrule
Linear Regression & 0.739 & 0.012 & 0.962 & 0.016 & 0.886 & 0.114 \\
Linear SVM & 0.709 & 0.034 & 1.004 & 0.046 & 0.877 & 0.123 \\
Decision Tree & 0.680 & 0.050 & 0.981 & 0.050 & 0.883 & 0.118 \\
Random Forest & 0.662 & 0.054 & 0.958 & 0.593 & 0.889 & 0.113 \\
XGBoost & 0.651 & 0.053 & 0.944 & 0.061 & 0.892 & 0.110 \\
LightGBM (All) & 0.591 & 0.014 & 0.857 & 0.014 & 0.912 & 0.088 \\
LightGBM (Best) & 0.590 & 0.013 & 0.857 & 0.013 & 0.912 & 0.088\\
\bottomrule
\end{tabular}
\label{tab:Table 2}
\end{table*}

For the LightGBM model, to prioritize the predictor variables based on their significance in predicting the rating, a sequential selection approach was adopted to achieve the best performance \cite{zhou2021using}. Starting from the most significant predictor variable, one less important variable was added to the model until the least important predictor was included. The performance of the sequentially selected models is shown in Fig.~\ref{fig:sequential_1}. As the number of features in the model increased, the MAE and RMSE decreased, and the correlation coefficient between predicted and true ratings increased. However, after introducing more features, the model's performance tended to stabilize. The optimal model performance was achieved by including the top 23 predictor variables.

\begin{figure}
     \centering
     \includegraphics[width=0.65\linewidth]{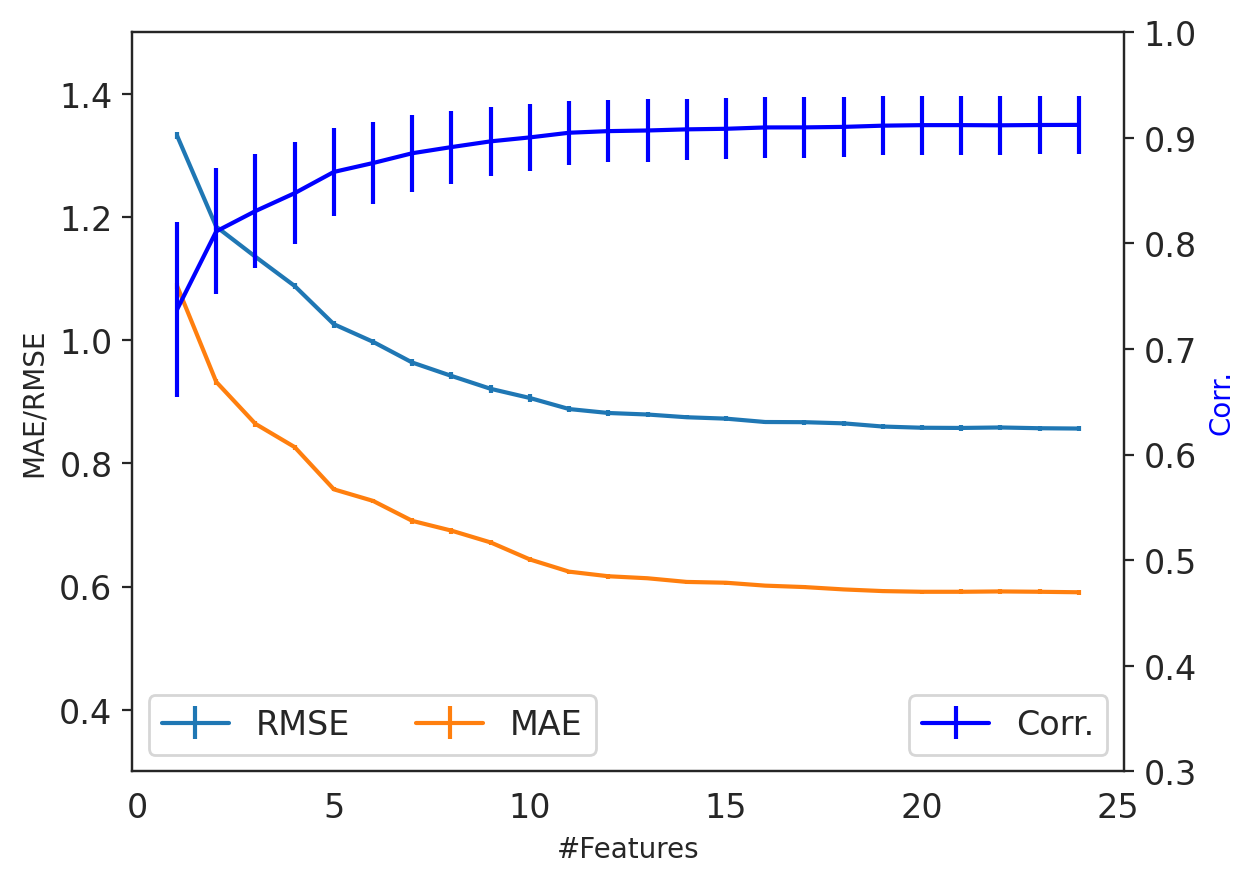}
     \caption{Performance of the micro-24 LightGBM models when predictor variables were sequentially selected from the most important to the less important ones}
     \label{fig:sequential_1}
\centering
\label{fig:sequential_1}
\end{figure}

\subsubsection{Macro-Level Model Performance}
The macro-level model was used to predict average station ratings. A set of regression models was compared with the LightGBM models, utilizing the 24 aspect-with-polarity predictor variables and the final model performed even better than the micro-level model, achieving a MAE of 0.483, an RMSE of 0.704, and a correlation coefficient of 0.920 as illustrated in Table~\ref{tab:Table 4}. The final model used a similar sequential selection method as described above with the top 22 features resulted in the best-performing LightGBM model (see Fig.~\ref{fig:sequential_3}). This favorable performance suggests that aggregating sentiment across multiple user reviews provides a more robust and accurate representation of overall station quality.

\begin{table*}[h] 
\centering
\caption{Performance of selected regression models with 24 aspect polarity predictor variables.}
\begin{tabular}{ccc|cc|cc} 
\toprule

Model & \multicolumn{2}{c|}{MAE} & \multicolumn{2}{c|}{RMSE} & \multicolumn{2}{c}{Corr.} \\ 
& Mean & SD & Mean & SD & Mean & SD \\ \midrule 

Linear Regression & 0.573& 0.028& 0.777& 0.052& 0.900& 0.102
\\
Linear SVM & 0.567& 0.029& 0.789& 0.055& 0.898& 0.105
\\
Decision Tree & 0.577& 0.033& 0.826& 0.077& 0.890& 0.113
\\
Random Forest & 0.557& 0.047& 0.805& 0.081& 0.895& 0.109
\\
XGBoost & 0.549& 0.048& 0.797& 0.079& 0.897& 0.106
\\
LightGBM (All) & 0.486& 0.026& 0.706& 0.035& 0.919& 0.082
\\
LightGBM (Best) & 0.483& 0.023& 0.704& 0.035& 0.920& 0.081
\\
\bottomrule
\end{tabular}
\label{tab:Table 4}
\end{table*}

\begin{figure}[h]
     \centering
     \includegraphics[width=0.65\linewidth]{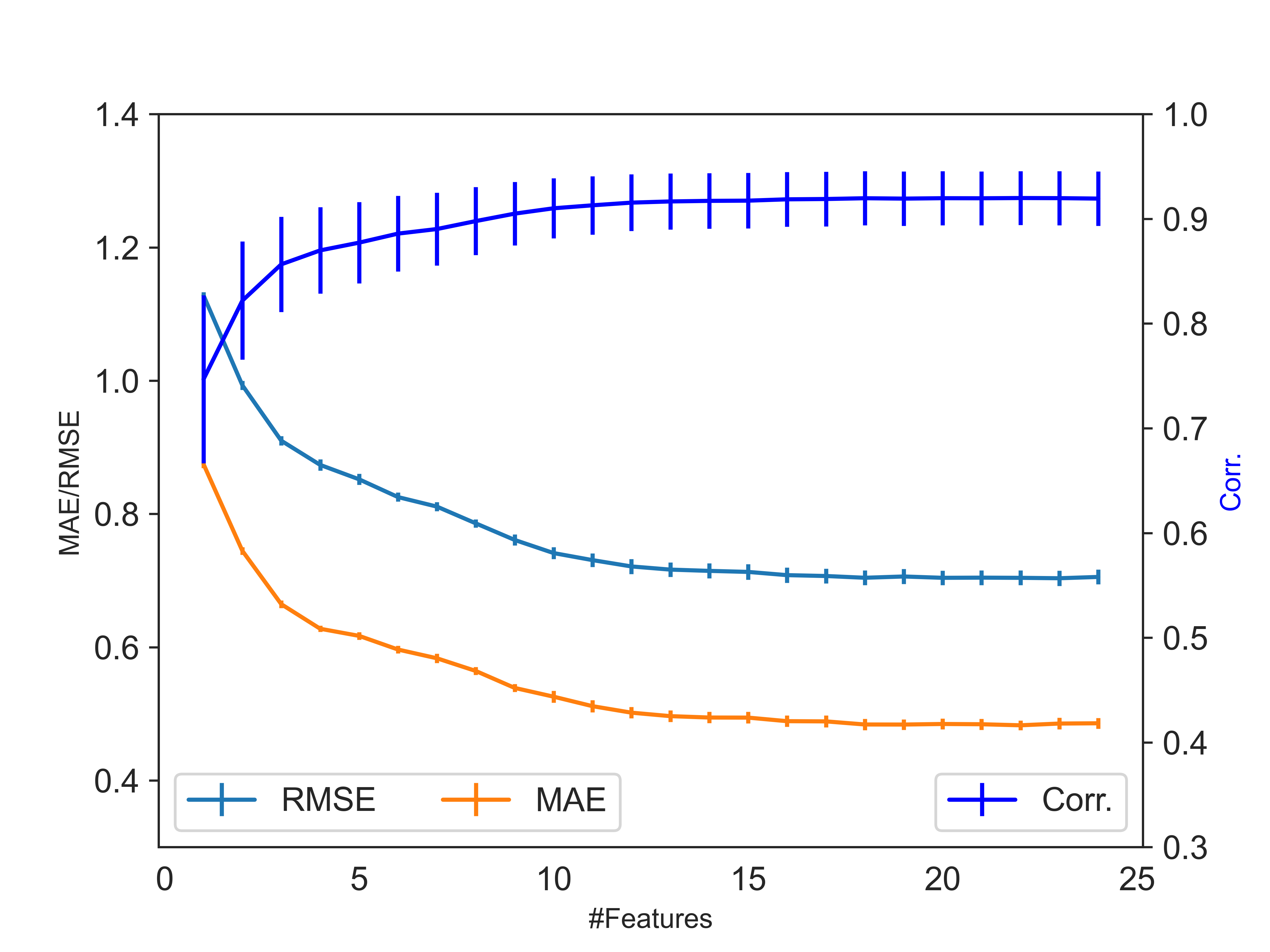}
     \caption{Performance of the macro 24 aspect with polarity LightGBM models with sequential selection}
     \label{fig:sequential_3}
\centering
\label{fig:sequential_3}
\end{figure}

\subsection{Interpreting Feature Importance with SHAP}

To gain a deeper understanding of the factors driving user satisfaction, we employed SHAP (SHapley Additive exPlanations) to analyze the LightGBM models. SHAP provides a game-theoretic approach to interpreting machine learning models, allowing us to quantify the contribution of each aspect to the model's predictions \cite{lundberg2020local}.

We computed SHAP values with a 10-fold cross-validation process to ensure that all samples in the dataset were considered for their influence on user ratings. The final SHAP values represent the average impact of each aspect across all folds. These values were then used to generate summary plots, which visualize the relative importance of each aspect in driving user satisfaction (See Fig.~\ref{fig:SHAP_Indiv} and Fig.\ref{fig:SHAP_avg}). In these plots, each dot represents a single user review, with its color indicating the presence or absence of the corresponding aspect and its position along the horizontal axis reflecting the magnitude and direction of the aspect's impact on the user rating. The vertical axis ranks the aspects based on their overall importance. 

\subsubsection{Micro-Level SHAP Analysis}
Based on the configuration of the data, all the predictor variables were binary for the micro-level model. Therefore, the blue dots (i.e., low value: 0) in the figure indicated that the specific aspect with polarity did not show up for the user reviews, while the red dots (i.e., high value: 1) indicated the presence of the aspect with the polarity identified in the review comments.

From Fig.~\ref{fig:SHAP_Indiv}, it is evident that the existence of positive aspects, such as ``amenities and location" (Positive), significantly enhanced user ratings. Conversely, negative aspects like ``reliability and maintenance" (Negative) negatively affected the ratings. For example, consider the first predictor variable, ``amenities and location\_Positive." A high value of predictor variable (i.e., 1) indicated that the presence of this aspect and polarity led to positive SHAP values, correlating with higher user ratings. Conversely, a low value (i.e., 0) indicating the absence of the amenity and location features was associated with a lower rating. The second most important aspect was ``reliability and maintenance\_Negative." The presence of this feature extended across the negative side of the SHAP values, suggesting an association with lower ratings, while the absence of the feature did not appear to have a significant impact, as most dots were clustered near the baseline of 0. The most significant aspects that impacted the customer’s ratings on the EV charging stations were ``amenities and location" (Positive), ``reliability and maintenance" (Negative), ``charging speed and efficiency" (Negative and Positive), ``accessibility and availability" (Negative and Positive). 

\begin{figure*}[h]
     \centering
     \includegraphics[width=0.65\linewidth]{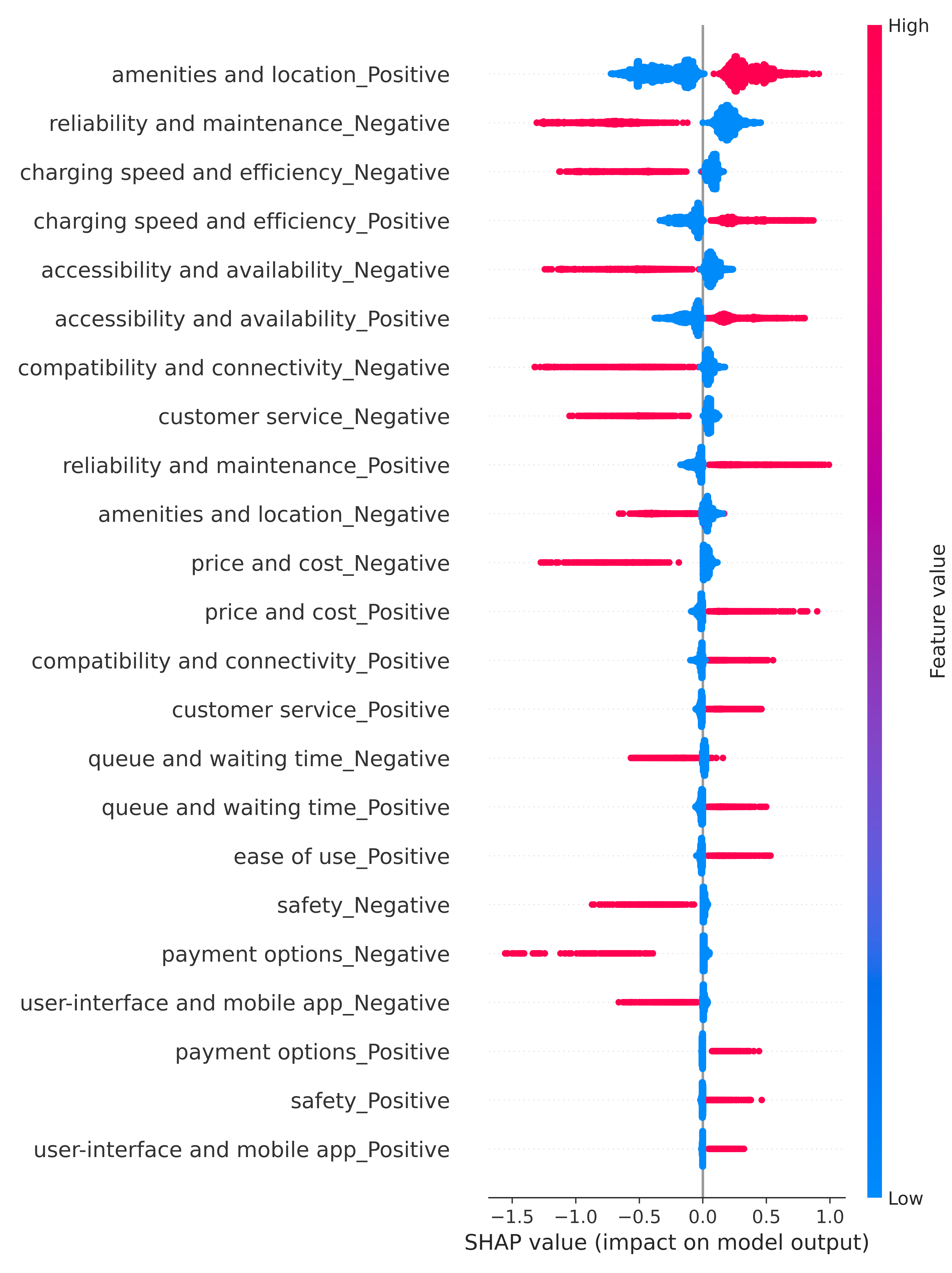}
     \caption{Importance ranking of the predictor variables in the Micro-level 24 variable LightGBM model (best) produced by SHAP}
     \label{fig:SHAP_Indiv}
\centering
\label{fig:SHAP_Indiv}
\end{figure*}

\begin{figure*}[t]
\begin{subfigure}[t]{0.5\textwidth} 
\vbox{
\centering{
  \includegraphics[width=1\linewidth]{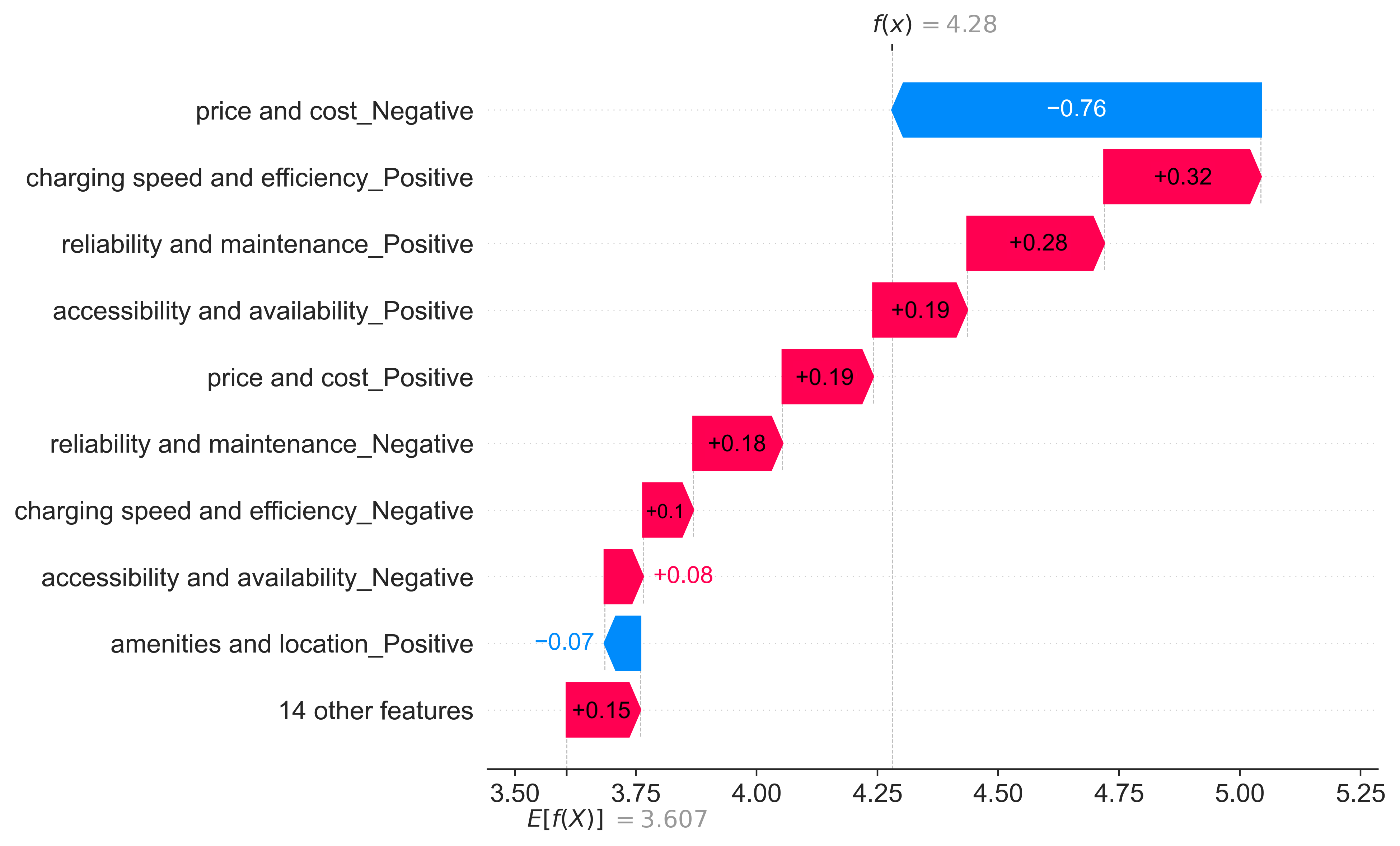}
}%
}%
\subcaption{\label{fig:sub1}}
\end{subfigure}%
\begin{subfigure}[t]{0.5\textwidth} 
\centering{
\includegraphics[width=1\linewidth]{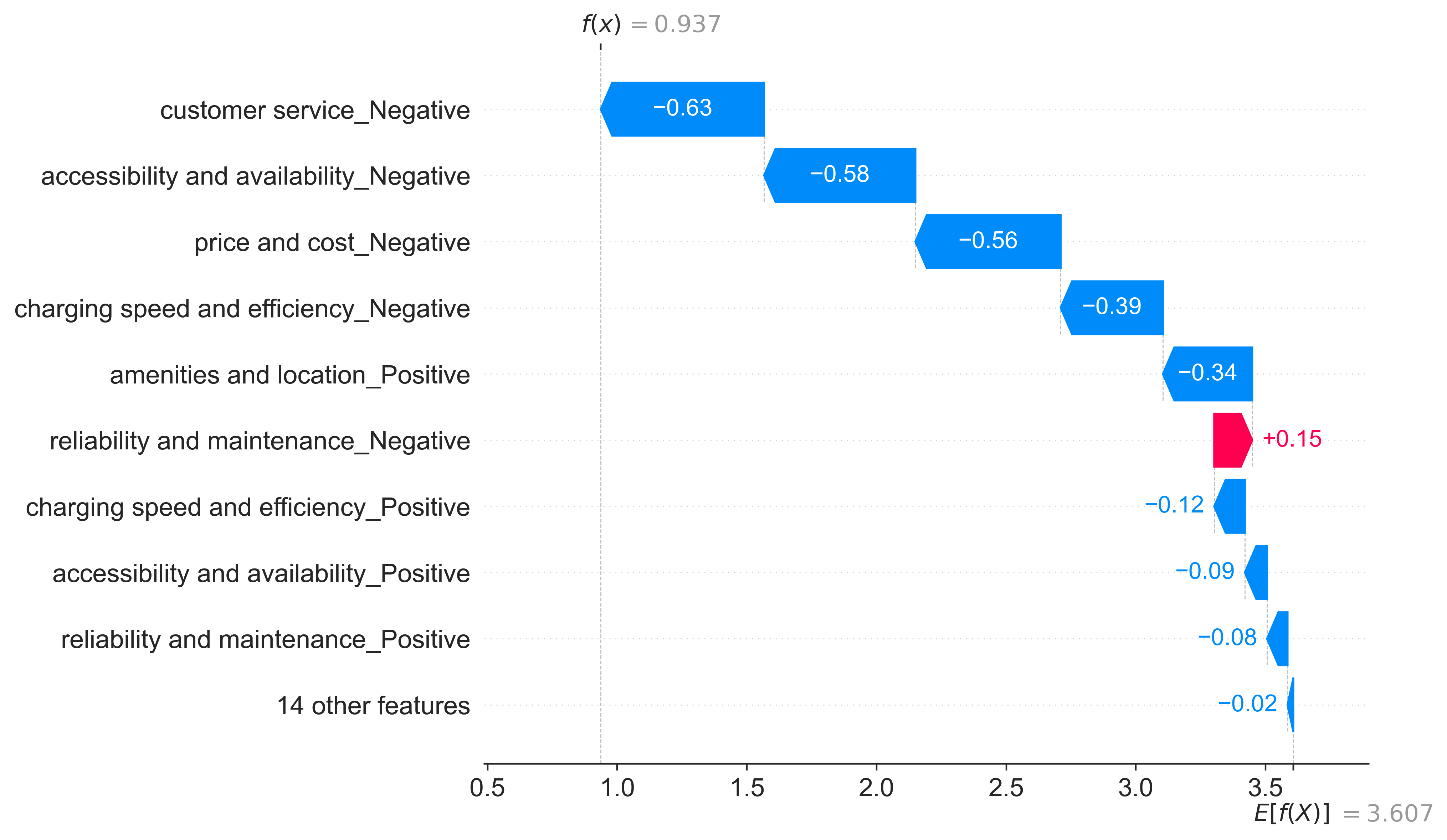}
\subcaption{\label{fig:sub2}}
}\end{subfigure}%
\caption{Individual instances SHAP values of LightGBM model (best) with: \subref{fig:sub1} high predicted rating and \subref{fig:sub2} low predicted rating \label{fig:6}}
\end{figure*}

The SHAP technique also provides insights into predicting individual instances within the dataset. Fig.~\ref{fig:6} illustrates two cases with contrasting predicted ratings, highlighting how different aspects influence user evaluations.

The baseline E[f(x)], representing the expected user rating in the dataset, is 3.607 (on a 0-5 scale). The bars with positive (red) and negative (blue) values indicated the increase and decrease from the predictor variables on the predicted user ratings for each sample. In the first instance (Fig.~\ref{fig:sub1}), the model predicted a high rating (f(x) = 4.28), which was close to the actual rating of 5. This user's positive experience was primarily driven by factors such as ``charging speed and efficiency" and ``reliability and maintenance," which positively contributed 0.32 and 0.28 to the rating, respectively. While ``price and cost" negatively impacted the rating (-0.76), its effect was offset by the positive aspects. At the bottom, the 14 less significant variables’ impact canceled out each other and produced a total increase of 0.15, which was less significant. For this user, though he/she experienced negative feelings on the price and cost of the charging station, the other positive aspects offset the effect and presented an overall positive feeling on the charging experience. The comment left by the customer was \textit{``It’s a free dc charger what can I say bad. It works and can do 110-120 kw if it’s just your car. They say they are going to start charging soon"}, highlighting the functionality and speed of the free DC charger as key factors outweighing the anticipated future cost.

Conversely, the second instance (Fig. \ref{fig:sub2}) represents a negative user experience with a low predicted rating (f(x) = 0.937) and an actual rating of 1. This dissatisfaction was largely attributed to ``customer service" (-0.63), ``accessibility and availability" (-0.58), ``price and cost" (-0.56), and ``charging speed and efficiency" (-0.39). The user's comment emphasized the exorbitant cost and slow charging speed as major sources of frustration: \textit{``Absurdly expensive. More expensive than pumping gas. I pay 5-10x the actual cost of electricity, and it’s the slowest charger I’ve ever used. I’m a captive consumer since this is my only charging option in my apartment complex. Shame on Blink for ripping people off like this. I don’t know how business practices like this are legal."}

These contrasting examples demonstrate how SHAP analysis can illuminate the complex interplay of factors that shape user satisfaction with EV charging stations, providing valuable insights for targeted improvements.

\subsubsection{Macro-Level SHAP Analysis}

For the charging station level analysis, 22 predictor variables were included in the best-performing LightGBM model. The SHAP values summary plot, illustrated in Fig.~\ref{fig:SHAP_avg}, offered insights into these variables. As the dataset was aggregated and averaged based on individual user reviews, the predictor values were represented on a continuous scale from 0 to 1. The color spectrum of the dots, ranging from blue to red, reflects the ascending values of the predictor variable. This color gradient signified the proportion of user reviews mentioning specific aspects with polarities towards a particular charging station. For example, if the predictor variable “amenities and location\_Positive” attained the value of 1 for a particular charging station, it meant that all the user reviews associated with that station had been identified as positive regarding its amenities and location. Conversely, if the predictor value was 0, none of the reviews for that station ever mentioned any positive comments regarding its amenities and location.

From Fig.~\ref{fig:SHAP_avg}, it can be noted that positive aspects significantly enhanced user ratings. Conversely, the presence of negative aspects imposed adverse effects on the ratings. For instance, the first predictor variable is ``amenities and location\_Positive." A higher value of the predictor variable, aggregated from more users' expressions of positive comments on the aspect, was correlated with higher user ratings. While a lower value indicating the absence of the amenity and accessibility features from users was associated with lower ratings. Neutral predictor variable values (i.e., purple points) representing charging stations with minimum mentions within this feature were concentrated in the middle area with minor SHAP values, indicating little to no impact on the user ratings. For the second aspect ``reliability and maintenance\_Negative", its presence was mostly observed on the negative side of the SHAP values, which were related to lower ratings. However, the absence of the feature did not demonstrate a notable impact, as the blue points were aggregated with minor positive SHAP values. The most significant aspects that impacted the customer’s ratings on the EV charging stations were amenities and location (Positive), reliability and maintenance (Negative), charging speed and efficiency (Negative and Positive), and accessibility and availability (Negative and Positive).  

\begin{figure}[h]
     \centering
     \includegraphics[width=0.65\linewidth]{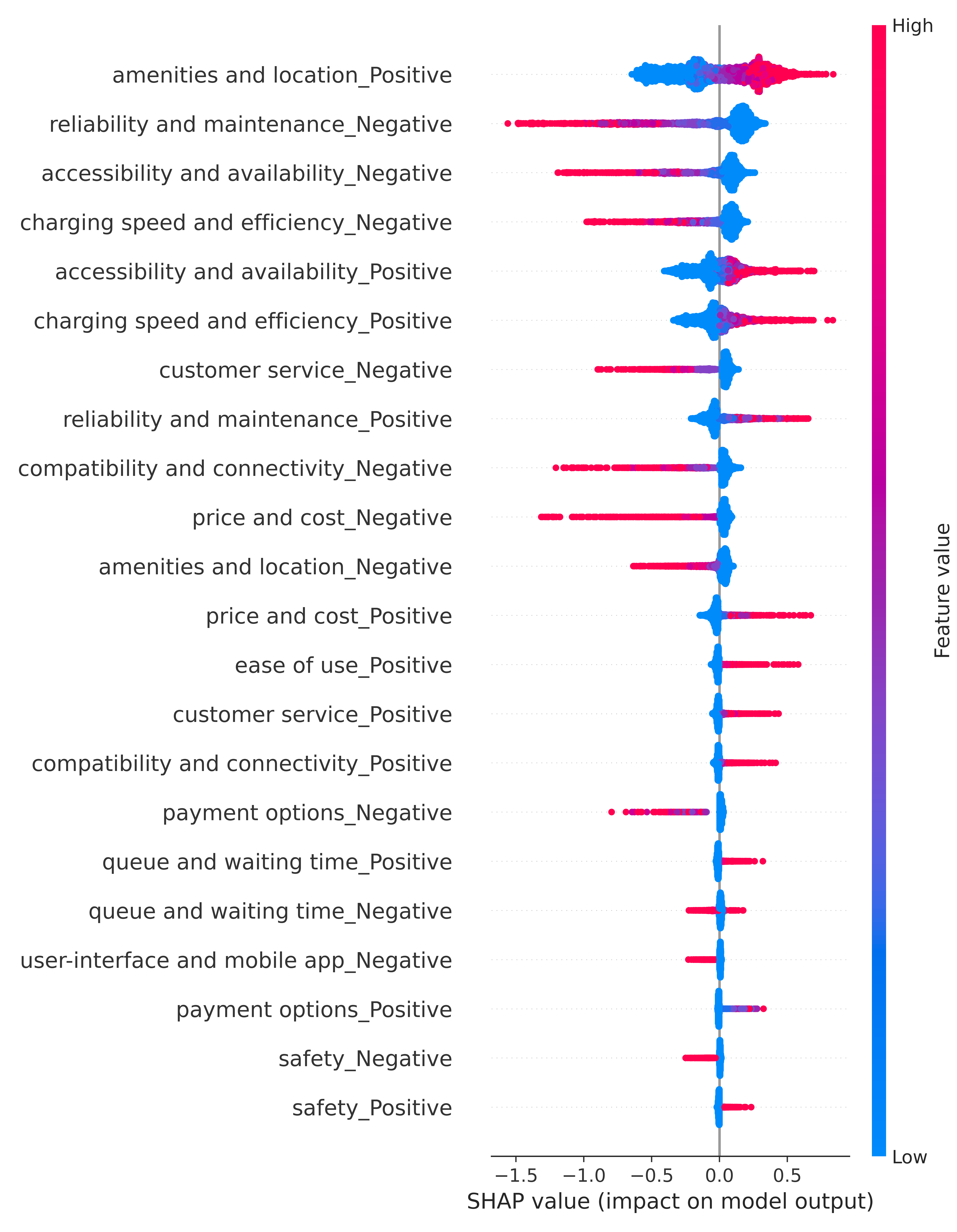}
     \caption{Importance ranking of the predictor variables in the Macro-level 24 variable LightGBM model (best) produced by SHAP
}
     \label{fig:SHAP_avg}
\centering
\label{fig:SHAP_avg}
\end{figure}

\subsubsection{Micro-Level vs. Macro-Level: Comparing Key Insights}

Our analysis revealed consistent trends in the significant factors influencing user ratings across both the micro-level (individual user) and macro-level (aggregated station) models. 

\textbf{Consistent Trends:} Both micro-level and macro-level models consistently identified positive sentiment towards amenities and location, along with negative sentiment regarding reliability and maintenance, as key drivers of user satisfaction. This suggests that these factors play a universal role in shaping user perceptions of charging stations.

\textbf{Differing Focus:} The error metrics for the macro-mode, including MAE (0.483) and RMSE (0.704), were both lower than the errors of the macro-level model (MAE: 0.590; RMSE: 0.857). Also, the correlation coefficient for the macro-level (0.920) was higher than the micro-level model (0.912), indicating a stronger correlation between the predicted ratings and the ground truth. Therefore, the macro-level model showed a slightly superior result compared to the micro-level model based on the evaluation metrics. By aggregating data from individual users, the macro-level analysis reduced the variability and noises introduced by outliers, thus showing a more robust analysis result.

While the top factors were consistent, their relative importance differed slightly between the two models. Individual users (micro-level) placed greater emphasis on charging speed and efficiency, while aggregated station ratings (macro-level) prioritized accessibility and availability. This difference highlights the need to consider both individual preferences and the overall availability of charging infrastructure when optimizing charging stations.

\textbf{Negative Sentiment Dominance:} Across both models, negative polarity had a stronger impact on user ratings than positive polarity. This suggests that addressing negative aspects, such as reliability and maintenance issues, should be a primary focus for charging station operators.

\section{Discussions}
In this study, we aimed to explore user preferences of electric vehicles when selecting charging stations, ultimately providing insights for mitigating range anxiety and enhancing user experiences within EV charging infrastructures.

\subsection{Comparison with Existing Literature}
Our findings align with existing literature on key aspects influencing EV user satisfaction, such as the importance of amenities, charging speed, and pricing \cite{ge2022charging, chen2017california, liu2018optimal, wang2021electric}. However, our study goes beyond previous research by leveraging large-scale real-world data and advanced machine learning techniques to provide a more comprehensive and nuanced understanding of user preferences. Notably, we highlight the critical role of charging station reliability and maintenance in shaping user satisfaction, an aspect often overlooked in prior studies.

While previous research often focused on objective factors influencing charging behavior \cite{morrissey2016future, jiang2021review}, our research emphasizes the subjective user experience and its impact on overall satisfaction. This highlights a crucial shift in focus towards user-centric design and operation of EV charging infrastructure.
\subsection{Addressing Research Gaps}

Our study addresses the limitations of existing approaches that primarily rely on surveys and questionnaires with limited sample sizes \cite{pevec2020survey, globisch2019consumer}. By leveraging a large dataset of Google Maps reviews, we captured diverse user experiences and preferences across a wider population. Moreover, our aspect-based sentiment analysis using ChatGPT 4.0 reveals fine-grained user sentiments towards specific aspects of charging stations, uncovering nuances that traditional methods might miss.

Despite the strengths of our approach, we acknowledge the inherent limitations of relying on user reviews. Online feedback platforms may introduce biases, and certain aspects of user experience might not be adequately captured in textual reviews. Furthermore, the performance of our sentiment analysis and preference models relies heavily on the quality and comprehensiveness of the training data.

\subsection{Implications for EV Industry}
Our findings provide actionable insights for various stakeholders in the EV industry:
\begin{itemize}
\item \textbf{Charging Station Operators:} Prioritize addressing reliability and maintenance concerns to enhance user satisfaction. Enhance location convenience by ensuring proximity to amenities like restaurants and shops. Provide real-time availability updates and transparent pricing information to address range anxiety.
\item \textbf{Policymakers:} Incentivize the development of charging infrastructure in high-demand areas identified through our analysis. Encourage the adoption of user-centric design principles and promote the inclusion of amenities at charging stations.
\item \textbf{EV Manufacturers:} Incorporate real-time availability information, predictive maintenance alerts, and user-friendly interfaces into vehicle dashboards to mitigate range anxiety and enhance the overall charging experience.
\end{itemize}

\subsection{Limitations and Future Directions}
Our study primarily focused on user reviews from Google Maps, which may not represent the entire EV user population. Future research could incorporate data from other sources like social media to gain a more holistic view of user sentiments. Moreover, our aspect-based analysis could be expanded to include additional dimensions, such as station cleanliness, customer support quality, and the impact of emerging technologies like wireless charging and vehicle-to-grid integration.
Future research could also investigate the influence of individual user characteristics, such as driving habits, personality traits, and demographics, on their charging preferences. Incorporating these factors into personalized recommendation systems could further enhance user satisfaction and encourage wider EV adoption.

\section{Conclusions}

In conclusion, this study delves into user preferences of electric vehicle (EV) users concerning charging stations, aiming to offer actionable insights into mitigating range anxiety concerns and elevating user experiences within EV charging infrastructures. Our research contributes to a deeper understanding of the factors influencing user satisfaction and preference through meticulous analysis of user sentiments and critical aspects of charging stations. By shedding light on these key elements, our findings pave the way for informed strategies to optimize EV charging station design, operation, and accessibility, ultimately fostering a more seamless and satisfying experience for EV users.

\section*{Acknowledgment} 

The authors would like to acknowledge Ford Motor Company for sponsoring and supporting this project. Any opinions, findings, and conclusions, or recommendations expressed in this paper are those of the authors and do not necessarily reflect the views of Ford.

\section*{Funding Data}
This research was supported by Ford Motor Company.





\bibliographystyle{unsrt}   

\bibliography{main} 



\end{document}